\documentclass[aps,pre,superscriptaddress,amsmath,preprint,showpacs]{revtex4}
\usepackage{amsmath}
\usepackage[dvips]{graphicx}

\newcommand{\be}{\begin{equation}}
\newcommand{\ee}{\end{equation}}

\newcommand{\fig}[2]{Figs  (\ref{#1}) \& (\ref{#2})} 
\newcommand{\Fig}[1]{Fig. (\ref{#1})} 
 \newcommand{\eqa}{\begin{eqnarray}}
\newcommand{\eeq}{\end{eqnarray}}  
\newcommand{\eqsto}[2]{Eqs. (\ref{#1}) to (\ref{#2})}

\begin{document}
\title{Statistics of magnetic field fluctuations in a partially
  ionized space plasma}

\author{Dastgeer Shaikh\footnote{\tt Email : dastgeer.shaikh@uah.edu}}
\affiliation{Department of Physics and 
Center for Space Plasma and Aeronomy Research, 
The University of Alabama in Huntsville, 
Huntsville, Alabama 35899, USA}

\begin{abstract}
{\em Voyager 1} and {\em 2} data reveals that magnetic field
fluctuations are compressive and exhibit a Gaussian distribution in
the compressed heliosheath plasma, whereas they follow a lognormal
distribution in a nearly incompressible supersonic solar wind
plasma. To describe the evolution of magnetic field, we develop a
nonlinear simulation model of a partially ionized plasma based on two
dimensional time-dependent multifluid model. Our model
self-consistently describes solar wind plasma ions, electrons,
neutrals and pickup ions.  It is found from our simulations that the
magnetic field evolution is governed by mode conversion process that
leads to the suppression of vortical modes, whereas the compressive
modes are amplified. An implication of the mode conversion process is
to quench the Alfv\'enic interactions associated with the vortical
motions. Consequently anisotropic cascades are reduced. This is
accompanied by the amplification of compressional modes that tend to
isotropize the plasma fluctuations and lead to a Gaussian distribution
of the magnetic field.

\end{abstract}
\pacs{52.25.Gj, 52.35.Fp, 52.50.Jm, 98.62.En}
\received{April 11,  2010}
\maketitle
\section{Introduction}

Space plasma is in a fully developed turbulent state
\cite{Shukla77,Shukla78}. Turbulent interactions are mediated by the
solar wind that emanates from the Sun and propagates outwardly. It
interacts with partially ionized interstellar gas predominantly via
charge exchange, and creates pick up ions \cite{zank,zank1996}.  Near
the termination shock (which is about 90-100AU from the Sun), the
supersonic solar wind decelerates, heats up, and it is compressed. It
becomes subsonic in a region called heliosheath.  In the heliosheath
region, the solar wind plasma is compressed.  The solar wind further
interacts with interstellar neutrals via charge exchange.  These
interactions are described comprehensively by Zank in
Ref. \cite{zank}.  During its journey from the Sun, the solar wind
plasma develops multitude of length and time scales that interact with
the partially ionized interstellar gas and nonlinear structures
develop in a complex manner.  Many features of the in situ heliosheath
plasma have been surprising and were not expected from the existing
analytic and simulation modeling. One of the most notable Voyager
observations is the solar wind plasma near the heliosheath is subsonic
and compressive \cite{bur}.  The subsonic and compressed solar wind
plasma exhibits a Gaussian distribution in magnetic field fluctuations
contrary to the lognormal that is typically observed in the non
compressive solar wind plasma \cite{bur}. The physical processes
leading to the Gaussian distribution in magnetic field fluctuations
are not understood.

A primary goal of this paper is to describe a self-consistent
evolution of the compressed solar wind plasma fluctuations by
examining why magnetic field fluctuations exhibit a Gaussian
distribution.  For this purpose, we develop a fully self-consistent
description of plasma-neutral coupled system and investigate
compressive and non-compressive characteristic of magnetic field
fluctuations in the context of partially ionized solar wind plasma.
This issue is critically important in space plasmas because of its
ramifications on origin of cosmic rays, energetic particles, partially
ionized turbulence and many other \cite{zank,zank1996,dastgeer}.

In section 2, we describe our new multi fluid model of plasma that is
coupled with neutral gas in a partially ionized environment. Our model
self-consistently describes the evolution of solar wind ions,
electrons, pickup ions and neutral fluids. Implicit in our model is
the interaction of small scale turbulence with a compressive plasma.
Section 3 describes our simulation results dealing with the
compressive characteristic of the solar wind plasma. Section 4
describes statistics of magnetic field fluctuations in compressive and
non-compressive MHD plasma and finally section 5 summarizes our major
findings.

\section{multifluid Turbulence Model}
Our nonlinear simulation model employs the dominant components of
multi fluid species of the solar wind plasma. It includes plasma
electrons, pickup ions, solar wind ions, and neutral gas. The solar
wind ions interact with the interstellar neutral hydrogen via charge
exchange that depends on the relative speeds of the solar wind and
neutral atoms \cite{zank,zank1996,dastgeer}. We assume that
fluctuations in the plasma and neutral fluids are isotropic,
homogeneous, thermally equilibrated and turbulent. The characteristic
turbulent correlation length-scales ($\lambda_c \sim 1/k_c$) are
typically bigger than charge-exchange mean free path lengths
($\lambda_{ce}\sim 1/k_{ce}$) in the space plasma flows, i.e
$\lambda_c\gg \lambda_{ce}$ or $k_{ce}/k_c \gg 1$.

The fluid model describing nonlinear turbulent
processes in the interstellar medium, in the presence of charge
exchange, can be cast into plasma density ($\rho_p$), velocity (${\bf
U}_p$), magnetic field (${\bf B}$), pressure ($P_p$) components
according to the conservative form
\be
\label{mhd}
 \frac{\partial {\bf F}_p}{\partial t} + \nabla \cdot {\bf Q}_p={\cal Q}_{p,n},
\ee
where,
\[{\bf F}_p=
\left[ 
\begin{array}{c}
\rho_p  \\
\rho_p {\bf U}_p  \\
{\bf B} \\
e_p
  \end{array}
\right], 
{\bf Q}_p=
\left[ 
\begin{array}{c}
\rho_p {\bf U}_p  \\
\rho_p {\bf U}_p {\bf U}_p+ \frac{P_p}{\gamma-1}+\frac{B^2}{8\pi}-{\bf B}{\bf B} \\
{\bf U}_p{\bf B} -{\bf B}{\bf U}_p\\
e_p{\bf U}_p
-{\bf B}({\bf U}_p \cdot {\bf B})
  \end{array}
\right],\\
{\cal Q}_{p,n}=
\left[ 
\begin{array}{c}
0  \\
{\bf Q}_M({\bf U}_p,{\bf V}_n, \rho_p, \rho_n, T_n, T_p)   \\
0 \\
Q_E({\bf U}_p,{\bf V}_n,\rho_p, \rho_n, T_n, T_p)
  \end{array}
\right]
\] 
and
\[ e_p=\frac{1}{2}\rho_p U_p^2 + \frac{P_p}{\gamma-1}+\frac{B^2}{8\pi}.\]
The above set of plasma equations is supplimented by $\nabla \cdot {\bf
B}=0$ and is coupled self-consistently to the  neutral density
($\rho_n$), velocity (${\bf V}_n$) and pressure ($P_n$) through a set
of hydrodynamic fluid equations,
\be
\label{hd}
 \frac{\partial {\bf F}_n}{\partial t} + \nabla \cdot {\bf Q}_n={\cal Q}_{n,p},
\ee
where,
\[{\bf F}_n=
\left[ 
\begin{array}{c}
\rho_n  \\
\rho_n {\bf U}_n  \\
e_n
  \end{array}
\right], 
{\bf Q}_n=
\left[ 
\begin{array}{c}
\rho_n {\bf U}_n  \\
\rho_n {\bf U}_n {\bf U}_n+ \frac{P_n}{\gamma-1} \\
e_n{\bf U}_n
  \end{array}
\right],\\
{\cal Q}_{n,p}=
\left[ 
\begin{array}{c}
0  \\
{\bf Q}_M({\bf V}_n,{\bf U}_p, \rho_p, \rho_n, T_n, T_p)   \\
Q_E({\bf V}_n,{\bf U}_p,\rho_p, \rho_n, T_n, T_p)
  \end{array}
\right],
\] 
\[e_n= \frac{1}{2}\rho_n V_n^2 + \frac{P_n}{\gamma-1}.\]

Equations (\ref{mhd}) to (\ref{hd}) form an entirely self-consistent
description of the coupled plasma-neutral turbulent fluid.  The
charge-exchange momentum sources in the plasma and the neutral fluids,
i.e. Eqs. (\ref{mhd}) and (\ref{hd}), are described respectively by
terms ${\bf Q}_M({\bf U}_p,{\bf V}_n,\rho_p, \rho_n, T_n, T_p)$ and
${\bf Q}_M({\bf V}_n,{\bf U}_p,\rho_p, \rho_n, T_n, T_p)$. A swapping
of the plasma and the neutral fluid velocities in this representation
corresponds, for instance, to momentum changes (i.e. gain or loss) in
the plasma fluid as a result of charge exchange with the neutral atoms
(i.e. ${\bf Q}_M({\bf U}_p,{\bf V}_n,\rho_p, \rho_n, T_n, T_p)$ in
Eq. (\ref{mhd})). Similarly, momentum change in the neutral fluid by
virtue of charge exchange with the plasma ions is indicated by ${\bf
  Q}_M({\bf V}_n,{\bf U}_p,\rho_p, \rho_n, T_n, T_p)$ in
Eq. (\ref{hd}). In the absence of charge exchange interactions, the
plasma and the neutral fluid are de-coupled trivially and behave as
ideal fluids.  While the charge-exchange interactions modify the
momentum and the energy of plasma and the neutral fluids, they
conserve density in both the fluids (since we neglect photoionization
and recombination). Nonetheless, the volume integrated energy and the
density of the entire coupled system will remain conserved in a
statistical manner. The conservation processes can however be altered
dramatically in the presence of any external forces. These can include
large-scale random driving of turbulence due to any external forces or
instabilities, supernova explosions, stellar winds, etc. Finally, the
magnetic field evolution is governed by the usual induction equation,
i.e. Eq. (\ref{mhd}), that obeys the frozen-in-field theorem unless
some nonlinear dissipative mechanism introduces small-scale damping.

Our model equations can be non-dimensionalized straightforwardly using
a typical scale-length ($\ell_0$), density ($\rho_0$) and velocity
($v_0$). The normalized plasma density, velocity, energy and the
magnetic field are respectively; $\bar{\rho}_p = \rho_p/\rho_0,
\bar{\bf U}_p={\bf U}_p/v_0, \bar{P}_p=P_p/\rho_0v_0^2, \bar{\bf
  B}={\bf B}/v_0\sqrt{\rho_0}$. The corresponding neutral fluid
quantities are $\bar{\rho}_n = \rho_n/\rho_0, \bar{\bf U}_n={\bf
  U}_n/v_0, \bar{P}_n=P_n/\rho_0v_0^2$. The momentum and the energy
charge-exchange terms, in the normalized form, are respectively
$\bar{\bf Q}_m={\bf Q}_m \ell_0/\rho_0v_0^2, \bar{Q}_e=Q_e
\ell_0/\rho_0v_0^3$. The non-dimensional temporal and spatial
length-scales are $\bar{t}=tv_0/\ell_0, \bar{\bf x}={\bf
  x}/\ell_0$. Note that we have removed bars from the set of
normalized coupled model equations (\ref{mhd}) \& (\ref{hd}).  The
charge-exchange cross-section parameter ($\sigma$), which does not
appear directly in the above set of equations, is normalized as
$\bar{\sigma}=n_0 \ell_0 \sigma$, where the factor $n_0\ell_0$ has
dimension of (area)$^{-1}$.  By defining $n_0, \ell_0$ through
$\sigma_{ce}=1/n_0\ell_0=k_{ce}^2$, we see that there exists a charge
exchange mode ($k_{ce}$) associated with the coupled plasma-neutral
turbulent system.  For a characteristic density, this corresponds
physically to an area defined by the charge exchange mode being equal
to (mpf)$^2$. The expressions for charge exchange sources are taken
from Refs \cite{zank, zank1996, dastgeer}.

\section{Simulation results}
We have developed a two-dimensional (2D) nonlinear fluid code to
numerically integrate \eqsto{mhd}{hd}.  The spatial discretization in
our code uses a discrete Fourier representation of turbulent
fluctuations based on a pseudospectral method, while we use a Runge
Kutta 4 method for the temporal integration. All the fluctuations are
initialized isotropically with random phases and amplitudes in Fourier
space. A mean ambient magnetic field is assumed to be present to
describe the large scale background magnetic field in the plasma. This
algorithm ensures conservation of total energy and mean fluid density
per unit time in the absence of charge exchange and external random
forcing. Additionally, $\nabla \cdot {\bf B}=0$ is satisfied at each
time step.  Our code is massively parallelized using Message Passing
Interface (MPI) libraries to facilitate higher resolution. The initial
isotropic turbulent spectrum of fluctuations is chosen to be close to
$k^{-2}$ with random phases in all three directions.  The choice of
such (or even a flatter than -2) spectrum does not influence the
dynamical evolution as the final state in our simulations progresses
towards fully developed turbulence.  While the turbulence code is
evolved with time steps resolved self-consistently by the coupled
fluid motions, the nonlinear interaction time scales associated with
the plasma $1/{\bf k} \cdot {\bf U}_p({\bf k})$ and the neutral
$1/{\bf k} \cdot {\bf V}_n({\bf k})$ fluids can obviously be
disparate. Accordingly, turbulent transport of energy in the plasma
and the neutral fluids takes place on distinctively separate time
scales.

\begin{figure}[ht]
\begin{center}
\includegraphics[width=8cm]{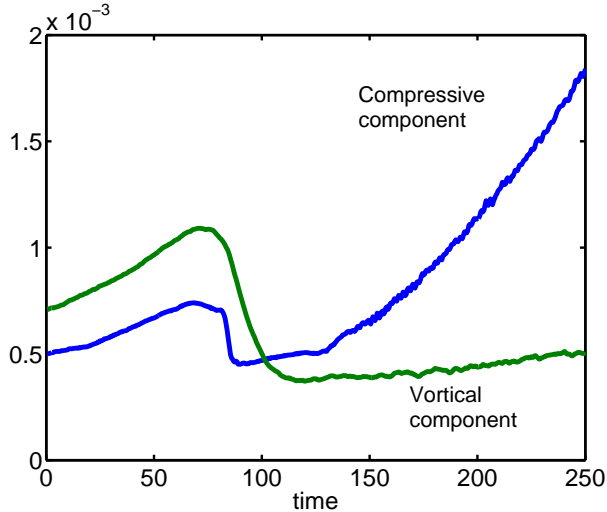}
\end{center}
\caption{\label{fig1} Quantitive evolution of compressive ($k_{com} =
  \sqrt{\sum_k |{\bf k} \cdot {\bf U}_p|^2/\sum_k |{\bf U}_p|^2}$) and
  vortical ($k_{vor}=\sqrt{\sum_k | {\bf k} \times {\bf U}_p|^2/\sum_k
    |{\bf U}_p|^2}$) components of plasma flow. Here ${\bf k}$ and
  $N_k$ are respectively the wave vector and total number of
  modes. The summation is carried over the entire turbulent spectrum.
  Initially, the vortical component is large. As time progresses,
  compressive component dominates over the vortical counter part in
  the heliosheath.}
\end{figure}

We now analyze statistics of magnetic field fluctuations in both
compressive and non-compressive MHD plasma to describe the statistics
of magnetic field fluctuations.  It should be noted that the initial
fluctuations in our simulations comprise both the vortical
(i.e. irrotational motion of fluid flow) and compressional (due to the
longitudinal flow motion) components. Our previous work show that the
vortical component of fluid flow dominates over the compressive
component in a supersonic solar wind plasma \cite{dastgeer2006}.  We
use this result as a basis to develop a self-consistent description of
compressive plasma fluctuations.  In the latter, the vortical motion
is sustained predominantly by shear Alfv\'enic modes that govern
nonlinear cascade in the solar wind plasma. The compressive modes, on
the other hand, are composed of fast/slow modes. The former survives,
whereas the latter decays in the solar wind \cite{dastgeer2006}.  By
contrast, the Voyager's observations indicate that the solar wind
becomes more compressive in the heliosheath plasma \cite{bur}. Hence
the compressive component of the flow is expected to dominate the
vortical component in the heliosheath plasma. To understand this
apparent discrepancy between the compressive and vortical modes in the
solar wind and heliosheath plasmas, we follow the evolution of the two
components in our simulations by initializing the velocity field with
a higher magnitude of vortical component. Our simulation results are
shown in \Fig{fig1} for $512^2$ modes in a two dimensional box of
length $2\pi \times 2\pi$. The other parameters in our simulations
are; charge exchange $k/k_{ce} \sim 0.1 - 0.01$, fixed time step
$dt=10^{-3}$, and collision parameter $\nu \sim 0.1 - 0.001$. The
background constant magnetic field $B_0=0.5$. Our simulations are
fully nonlinear because the ratio of the mean and fluctuating magnetic
fields $\delta {\bf B}/{\bf B}_0 \sim 1$.

As the evolution proceeds, nonlinear interactions quench the vortical
component and amplifies the compressive counterpart. Consequently, the
latter grows, while the former decays eventually and stays constant.
The compression of the velocity field corresponds essentially to the
compression of the magnetic field by virtue of the field and flow that
are coupled strongly under the ideal frozen-in-field state.  Our multi
fluid simulations thus demonstrate that progressive development of
compressive turbulence plays a catalyzing role in the mode coversion
(vortical to compressive) process.  Once the mode conversion process
is over, the two components decouple permanently and evolve
independent of each other. Further growth of the compressive component
in our simulation is ascribed to turbulent fluctuations that are
converted predominantly into the compressive mode by nonlinear
processes whereas minimal or almost no flux of energy is transmitted
into the vortical motion.  Our simulations results, describing the
predominance of the compressive modes over the vortical as shown in
\Fig{fig1}, are qualitatively consistent with the Voyager's
observations \cite{bur}.

\begin{figure}[t]
\begin{center}
\includegraphics[width=8.0cm]{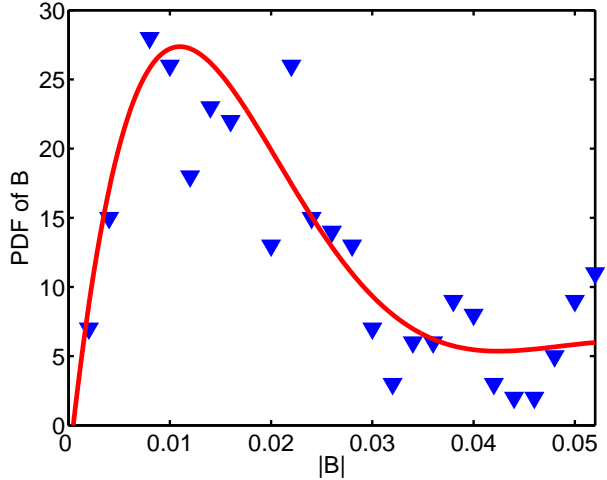}
\end{center}
\caption{\label{fig2} The magnetic field fluctuations in the
  non-compressive MHD turbulence exhibits a lognormal distribution in
  our simulations.  The magnetic field fluctuations are dominated by
  the vortical modes.}
\end{figure}

\section{statistics of magnetic field fluctuations}
\begin{figure}[t]
\begin{center}
\includegraphics[width=8cm]{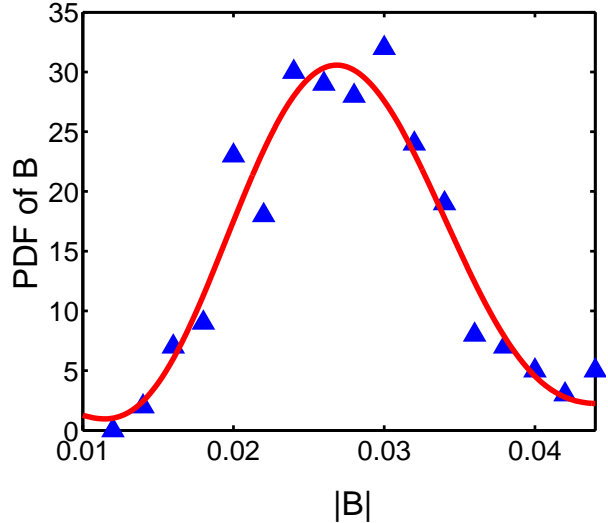}
\caption{\label{fig3} Probability distribution function of the
  magnetic field in the post-shock is consistent with the Gaussian
  distribution. The magnetic field fluctuations are dominated by the
  compressive modes.}
\end{center}
\end{figure}

We next analyze the magnetic field fluctuations in our
simulations. The results of our simulations, shown in
\fig{fig2}{fig3}, describe probability distribution function (PDF) of
the magnetic field fluctuations respectively in the non-compressive
and compressive regimes of solar wind MHD turbulence.  The PDF of the
magnetic field in the non-compressive plasma is consistent with the
lognormal distribution. This is shown in \Fig{fig2}. By contrast, the
magnetic field in the compressive plasma follows a Gaussian PDF as
shown in \Fig{fig3}. The latter is consistent with the Voyager 1
observations as reported by Burlaga et al. \cite{bur}. A Gaussian PDF
corresponds typically to a uniform, random and isotropic distribution,
and a mean deviation in any of the latter leads essentially to a
skewed or lognormal distribution \cite{Limpert}. In the context of our
simulations [see \fig{fig1}{fig2}], we infer that a lognormal
distribution of $B$ in the non-compressive plasma results primarily by
the predominance of vortical motion in magnetized plasma that
primarily gives rise to Alfv\'enic-like fluctuations. In the presence
of a mean or background magnetic field, Alfv\'enic fluctuations tend
to anisotropize the energy cascades \cite{kr,sheb}. Consequently,
migration of turbulent energy is non symmetric along and across the
mean magnetic field.  The anisotropic cascade is therefore a process
that could potentially lead to a skewed or lognormal distribution of
the magnetic field in the non-compressive region which is dominated by
nearly incompressible vortical motion. By contrast, plasma is
dominated by the high frequency fluctuations in the compressed
region. The effect of Alfv\'en waves is relatively weak in this region
as compared to the vortically dominated non-compressive plasma.  Owing
thus to the weaker Alfv\'enic effect, the compressional and relatively
high frequency motions in plasma tend to isotropize the magnetic field
fluctuations. Hence a Gaussian PDF follows in the compressive plasma.
The physical process, describing how Gaussian and lognormal
distributions occur respectively in compressive and non-compressive
plasma, is consistent with our simulations shown in \fig{fig1}{fig2}.

\section{Conclusions}
In summary, we have investigated evolution of the magnetic field
fluctuations in small scale compressive and non-compressive MHD plasma
turbulence in a partially ionized environment.  Our results are useful
in describing the magnetic field data from Voyagers \cite{bur}.  We
find that initial turbulent fluctuations, comprising both the vortical
and compressive motion, evolve towards a state in which the vortical
motion predominantly governs nonlinear interaction in the
non-compressive plasma by exciting Alfv\'enic modes. By contrast, the
mode conversion process in the compressive plasma leads to the
suppression of shear Alfv\'enic vortical modes whereas the compressive
modes are amplified. The latter isotropizes the PDF of magnetic field
in the compressive plasma.

To summarize our findings, we find that the probability distribution
function of magnetic field in compressive MHD fluctuations is a
Gaussian.  The depleted vortical motions suppress the Alfv\'enic modes
in the compressive MHD plasma. This we believe is one of the
plaussible reasons why the magnetic field fluctuations are transformed
into a Gaussian (from the lognormal) in partially ionized compressive
solar wind plasma turbulence.  Our results, consistent with the
Voyager observations \cite{bur} and theoretical predictions
\cite{zank,zank1996}, may be useful in the context of heliospheric
plasma where charge exchange interactions govern numerous features of
the solar wind plasma \cite{zank,bur,dastgeer}.

\section*{Acknowledgments}
The partial support of NASA grants NNX09AB40G, NNX07AH18G, NNG05EC85C,
NNX09AG63G, NNX08AJ21G, NNX09AB24G, NNX09AG29G, and NNX09AG62G.  is
acknowledged.

\end{document}